\documentclass[aps,preprint]{revtex4}%
\usepackage{amsfonts}
\usepackage{amsmath}
\usepackage{amssymb}
\usepackage{graphicx}%
\begin{document}
\title{{\LARGE SHEAR \ AND \ VORTICITY \ IN \ AN \ ACCELERATING \ BRANS-DICKE
\ LAMBDA-UNIVERSE \ WITH \ TORSION}}
\author{Marcelo Samuel Berman$^{1}$}
\affiliation{$^{1}$Instituto Albert Einstein / Latinamerica\ - Av. Candido Hartmann, 575 -
\ \# 17}
\affiliation{80730-440 - Curitiba - PR - Brazil - email: msberman@institutoalberteinstein.org}
\date{21 June, 2008.}

\begin{abstract}
We study accelerating Universes with power-law scale-factors. We include shear
and vorticity, a cosmological "constant" term, and spin from torsion, as in
Einstein-Cartan's theory when a scalar-field of Brans-Dicke type acts in the
model. We find a "no-hair" result, for shear and vorticity; we also make
contact with the alternative Machian picture of the Universe.


\end{abstract}
\maketitle

\begin{center}
\bigskip{\LARGE SHEAR \ AND \ VORTICITY \ IN \ AN \ ACCELERATING \ BRANS-DICKE
\ LAMBDA-UNIVERSE \ WITH \ TORSION}

\bigskip Marcelo Samuel Berman
\end{center}

{\LARGE \bigskip I. Introduction}

\bigskip The existence of shear ( $\sigma$\ ), vorticity ( $\varpi$\ ) and
cosmological "constant" ( $\Lambda$\ ), have not been well discussed in the
context of a scalar field theory, say, compatible with Brans-Dicke (1961)
theory, with or without torsion, for an accelerating present Universe. The
purpose of this paper is two-fold:

i) to show that there is a "no-hair" power-law result, under which shear and
vorticity are erased from the model as time passes by; and to describe a valid model.

ii) to show that when the space is torsioned by a spin of the Universe, the
most obvious model, results in the same kind of time-evolution for the spin as
in the Machian Universe, as described by Berman(2007; 2007a, b, c; 2008;
2008a, b), thus pointing to the rotation, plus expansion, state.

\bigskip

Prior work by Berman (references above) on an inflationary solution for the
exponential expansion of the same kind of lambda-Universe, resulted in the
same conclusions.

\bigskip

We first analyse a Brans-Dicke scalar-field (Section II and III), and later
treat the inclusion of torsion \textit{\`{a} la }Einstein-Cartan, while
keeping the scalar-field.

\bigskip

{\LARGE II. Brans-Dicke model with shear and vorticity}

\bigskip

From the scalar-tensor cosmologies, the Brans-Dicke theory renders the most
simple case (Brans and Dicke, 1961). For details on this, and other
scalar-tensor theories, we advise the reader to consult the books by Berman
(2007a), Faraoni (2004) and Fujii and Maeda (2003).

By the same procedure as in the inflationary case, we take Raychaudhuri's
equation for this case, in \textit{Einstein's frame}, which mimicks Einstein's
theory in non-conventional units (Dicke, 1962), whereby the scale-factor and
time are scaled by \ \ $\phi^{-\frac{1}{2}}$\ \ \ \ while\ mass goes with
\ \ $\phi^{\frac{1}{2}}$\ \ . Afterwards, we shall proceed to the conversion
to the \textit{Jordan-frame}, i.e., in the conventional set of units.

\bigskip

Consider now\ \ Robertson-Walker's metric, \ \ \ 

\bigskip

$ds^{2}=dt^{2}-\frac{R^{2}(t)}{\left[  1+\left(  \frac{kr^{2}}{4}\right)
\right]  ^{2}}d\sigma^{2}$ \ \ \ \ \ \ \ \ \ \ \ \ \ \ , \ \ \ \ \ \ \ \ \ \ \ \ \ \ \ \ \ \ \ \ \ \ \ \ \ \ \ \ \ \ \ \ \ \ \ \ \ \ \ \ \ \ \ (1)

\bigskip

where \ \ $k=0$\ \ and \ $d\sigma^{2}=dx^{2}+dy^{2}+dz^{2}$\ \ .

\bigskip

\bigskip When we include a cosmological "constant" term, the field equations
in the \textit{Einstein's frame} read, for a perfect fluid,

\bigskip

$\frac{8\pi G}{3}\left(  \rho+\frac{\Lambda}{\kappa}+\rho_{\lambda}\right)
=H^{2}\equiv\left(  \frac{\dot{R}}{R}\right)  ^{2}$ \ \ \ \ \ \ \ \ \ \ \ . \ \ \ \ \ \ \ \ \ \ \ \ \ \ \ \ \ \ \ \ \ \ \ \ \ \ \ \ \ \ \ \ \ \ \ \ \ \ (2)

\bigskip

$-8\pi G\left(  p-\frac{\Lambda}{\kappa}+\rho_{\lambda}\right)  =H^{2}%
+\frac{2\ddot{R}}{R}$ \ \ \ \ \ \ \ \ \ \ \ \ \ \ \ \ . \ \ \ \ \ \ \ \ \ \ \ \ \ \ \ \ \ \ \ \ \ \ \ \ \ \ \ \ \ \ \ \ \ \ \ (3)

\bigskip

In the above, we have: \ 

$\bigskip$

$\rho_{\lambda}=\frac{2\omega+3}{32\pi G}\left(  \frac{\dot{\phi}}{\phi
}\right)  ^{2}=\rho_{\lambda0}\left(  \frac{\dot{\phi}}{\phi}\right)  ^{2}$
\ \ \ \ \ \ \ . \ \ \ \ \ \ \ \ \ \ \ \ \ \ \ \ \ \ \ \ \ \ \ \ \ \ \ \ \ \ \ \ \ \ \ \ \ \ \ \ \ \ \ \ \ \ \ (4)

\bigskip

\bigskip The complete set of field equations, must be complemented by three
more equations, of which only two are independent, when taken along with
\ (2),(3) and (4): one is the dynamical fluid equation; the other is the
continuity one, as follow:

\bigskip

$\frac{d}{dt}\left[  \left(  \rho+\rho_{\lambda}+\frac{\Lambda}{\kappa
}\right)  R^{3}\right]  +3R^{2}\dot{R}\left[  p-\frac{\Lambda}{\kappa}%
+\rho_{\lambda}\right]  =0$\ \ \ \ \ \ \ \ , \ \ \ \ \ \ \ \ \ \ \ \ \ \ \ \ \ \ \ \ \ \ \ (5)

\bigskip

and,

\bigskip

$\frac{d}{dt}\left[  \left(  \rho+\frac{\Lambda}{\kappa}\right)  R^{3}\right]
+3R^{2}\dot{R}\left[  p-\frac{\Lambda}{\kappa}\right]  +\frac{1}{2}R^{3}%
\frac{\dot{\phi}}{\phi}\left[  \rho+\frac{4\Lambda}{\kappa}-3p\right]
=0$\ \ \ \ \ \ \ . \ \ \ \ \ (6)

\bigskip

\bigskip Raychaudhuri's equation (Raychaudhuri, 1979), \ as derived by Berman,
becomes, when we include shear and vorticity, 

\bigskip

$3\dot{H}+3H^{2}=\dot{u}_{;\alpha}^{\alpha}+2\left(  \varpi^{2}-\sigma
^{2}\right)  -4\pi G\left(  \rho+3p+4\rho_{\lambda}\right)  +\Lambda$
\ \ \ \ \ \ \ \ , \ \ \ \ \ \ \ (7)\ \ \ 

\bigskip

\bigskip where, \ \ $H$\ \ \ and \ \ \ $\dot{u}_{;\alpha}^{\alpha}$\ \ stand
for Hubble's parameter, and the acceleration of the fluid, which from now on,
will be considered null ( $\dot{u}_{;\alpha}^{\alpha}=0$\ ).\ 

\bigskip

We remember that we first deal with a kind of Einstein's cosmology, so that
\ $\Lambda$\ \ stands constant and we make later the transformation towards
conventional units or, Jordan's frame. In the next Section, we present a
power-law model that encompasses the possible accelerating Universe, i.e.,
where the deceleration parameter is negative (\ $q<0$\ ).

\bigskip

{\LARGE III. Accelerating power-law BD-Universe}

\bigskip

Consider power-law scale-factors, with constant deceleration parameter, like
in the original papers by Berman (1983) and Berman and Gomide (1988).

\bigskip$R=(mDt)^{\frac{1}{m}}$ \ \ \ \ \ \ \ \ \ \ \ \ \ \ \ \ \ ,
\ \ \ \ \ \ \ \ \ \ (\ \ $D=$ \ constant \ ) \ \ \ \ \ \ \ \ \ \ \ \ \ \ \ \ \ \ \ \ (8)

\bigskip\bigskip where,

$m=q+1=$ \ constant \ \ \ \ . \ \ \ \ \ \ \ \ \ \ \ \ \ \ \ \ \ \ \ \ \ \ \ \ \ \ \ \ \ \ \ \ \ \ \ \ \ \ \ \ \ \ \ \ \ \ \ \ \ \ \ \ \ \ \ \ \ \ (9)

\bigskip

We now find the following solution,

\bigskip

$H=\left(  mt\right)  ^{-1}$ \ \ \ \ \ \ \ \ \ \ \ \ \ \ \ \ ; \ \ \ \ \ \ \ \ 

\bigskip

$\Lambda=$ $\Lambda_{0}t^{-2}$ \ \ \ \ \ \ \ \ \ \ \ \ \ \ \ \ \ \ \ ;

\bigskip

$\varpi^{2}=\varpi_{0}^{2}t^{\gamma}$ \ \ \ \ \ \ \ \ \ \ \ \ \ \ \ \ \ \ ;

\bigskip

$\sigma^{2}=\sigma_{0}^{2}t^{\gamma}$
\ \ \ \ \ \ \ \ \ \ \ \ \ \ \ \ \ \ \ \ ; \ \ \ \ \ \ \ \ \ \ \ \ \ \ \ \ \ \ \ \ \ \ \ \ \ \ \ \ \ \ \ \ \ \ \ \ \ \ \ \ \ \ \ \ \ \ \ \ \ \ \ \ \ \ \ \ \ \ \ \ (10)

\bigskip

$\rho=\rho_{0}t^{\beta}$ \ \ \ \ \ \ \ \ \ \ \ \ \ \ \ \ \ \ \ \ ;

\bigskip

$p=p_{0}t^{\beta}$ \ \ \ \ \ \ \ \ \ \ \ \ \ \ \ \ \ \ \ \ ;

\bigskip

$\rho_{\lambda}=\delta^{2}\rho_{\lambda0}t^{-2}$ \ \ \ \ \ \ \ \ \ \ \ \ \ ;

\bigskip

\bigskip and,

$\bigskip$

$\phi=\phi_{0}t^{\delta}$ \ \ \ \ \ \ \ \ \ \ \ \ \ \ \ \ \ \ \ \ \ ,

\bigskip

\bigskip where, \ \ \ $\Lambda_{0}$ ,\ \ $\varpi_{0}^{2}$\ , \ \ $\sigma
_{0}^{2}$\ \ ,\ \ \ $\rho_{0}$\ , \ \ $p_{0}$\ \ ,\ \ \ $\rho_{\lambda0}$\ \ ,
$\beta$\ \ , \ \ $\gamma$\ \ \ , \ $\delta$\ \ and\ \ \ $\phi_{0}$\ \ \ \ are constants.

\bigskip

In order that the model be consistent, we need to impose the following conditions,

\bigskip

$\beta=\gamma=-2$ \ \ \ \ \ \ \ \ \ \ \ \ \ \ \ \ \ \ \ \ , \ \ \ \ \ \ \ \ \ \ \ \ \ \ \ \ \ \ \ \ \ \ \ \ \ \ \ \ \ \ \ \ \ \ \ \ \ \ \ \ \ \ \ \ \ \ \ \ \ \ \ \ \ \ \ \ \ (11)

\bigskip

$\frac{3}{m}\left(  \frac{1}{m}-1\right)  =2\left(  \varpi_{0}^{2}-\sigma
_{0}^{2}\right)  -4\pi\left(  \rho_{0}+3p_{0}+4\delta^{2}\rho_{\lambda
0}\right)  +\Lambda_{0}$ \ \ \ \ \ .\ \ \ \ \ \ (12)

\bigskip

When we go back to Jordan's frame, we find:

\bigskip

$\bar{R}=R\phi^{\frac{1}{2}}$ \ \ \ \ \ \ \ \ ; \ \ \ \ \ \ \ \ \ \ \ \ \ \ \ \ \ \ \ \ \ \ \ \ \ \ \ \ \ \ \ \ \ \ \ \ \ \ \ \ \ \ \ \ \ \ \ \ \ \ \ \ \ \ \ 

\bigskip

$\bar{\rho}=\rho\phi^{-2}$\ \ \ \ \ \ \ \ ;

\bigskip

$\bar{p}=p\phi^{-2}=\left[  \frac{p_{0}}{\rho_{0}}\right]  \bar{\rho}$\ \ \ \ \ \ \ \ \ ;

\bigskip\ \ \ \ \ \ \ \ \ \ \ \ \ \ \ \ \ \ \ \ \ \ \ \ \ \ \ \ \ \ \ \ \ \ \ \ \ \ \ \ \ \ \ \ \ \ \ \ \ \ \ \ \ \ \ \ \ \ \ \ \ \ \ \ \ \ \ \ \ \ \ \ \ \ \ \ \ \ \ \ \ \ \ \ \ \ \ \ \ \ \ \ \ \ \ \ \ \ \ \ \ (13)

$\bar{\varpi}=\varpi\phi^{-\frac{1}{2}}$\ \ \ \ \ \ \ \ \ \ \ \ \ \ ;

\bigskip

$\bar{\sigma}=\sigma\phi^{-\frac{1}{2}}$ \ \ \ \ \ \ \ \ \ \ \ \ \ \ ;

\bigskip

$\bar{\Lambda}=\Lambda_{0}\phi^{-1}$\ \ \ \ \ \ \ \ \ \ \ \ \ ;

\bigskip

\bigskip$\bar{\phi}=\phi$\ \ \ \ \ \ \ \ \ \ \ \ \ \ \ \ \ \ \ \ ;

\bigskip

$\bar{H}=H\phi^{-\frac{1}{2}}$ \ \ \ \ \ \ \ \ \ \ \ \ \ \ \ \ \ \ \ \ \ \ \ .

\bigskip

\bigskip When we plug (10) and (11) in (13), we obtain,

\bigskip

$\bar{R}=\phi_{0}^{\frac{1}{2}}\left(  mD\right)  ^{\frac{1}{m}}t^{\left(
\frac{\delta}{2}+\frac{1}{m}\right)  }$ \ \ \ \ \ \ \ \ ; \ \ \ \ \ \ \ \ \ \ \ \ \ \ \ \ \ \ \ \ \ \ \ \ \ \ \ \ \ \ \ \ \ \ \ \ \ \ \ \ \ \ \ \ \ \ \ \ \ \ \ \ \ \ \ 

\bigskip

$\bar{\rho}=\rho_{0}\phi_{0}^{-2}t^{-2\left(  \delta+1\right)  }$\ \ \ \ \ \ \ \ \ \ \ \ \ \ \ \ \ \ ;

\bigskip

$\bar{p}=p_{0}\phi_{0}^{-2}t^{-2\left(  \delta+1\right)  }$\ \ \ \ \ \ \ \ \ \ \ \ \ \ \ \ \ \ ;

\bigskip\ \ \ \ \ \ \ \ \ \ \ \ \ \ \ \ \ \ \ \ \ \ \ \ \ \ \ \ \ \ \ \ \ \ \ \ \ \ \ \ \ \ \ \ \ \ \ \ \ \ \ \ \ \ \ \ \ \ \ \ \ \ \ \ \ \ \ \ \ \ \ \ \ \ \ \ \ \ \ \ \ \ \ \ \ \ \ \ \ \ \ \ \ \ \ \ \ \ \ \ (14)

$\bar{\varpi}=\varpi_{0}\phi_{0}^{-\frac{1}{2}}t^{-\left(  1+\frac{\delta}%
{2}\right)  }$\ \ \ \ \ \ \ \ \ \ \ \ \ \ ;

\bigskip

$\bar{\sigma}=\sigma_{0}\phi_{0}^{-\frac{1}{2}}t^{-\left(  1+\frac{\delta}%
{2}\right)  }$\ \ \ \ \ \ \ \ \ \ \ \ \ \ \ ;

\bigskip

$\bar{\Lambda}=\Lambda_{0}\phi_{0}^{-1}t^{-\left(  \delta+2\right)  }$\ \ \ \ \ \ \ \ \ \ \ \ \ \ \ \ \ \ ;

\bigskip

$\bar{H}=m^{-1}\phi_{0}^{-\frac{1}{2}}t^{-\left(  1+\frac{\delta}{2}\right)
}$ \ \ \ \ \ \ \ \ \ \ \ .

\bigskip

\bigskip It can be checked that we have the "no-hair" condition, for an
accelerating Universe, with \ $\delta>-1$\ \ and also \ $\delta>-\frac{2}{m}%
$\ \ , the scale-factor increases up to infinity when we shall find that
\ Hubble's parameter \ \ $\bar{H}\longrightarrow0$\ \ \ . The condition
\ \ $\delta>-2$\ \ \ \ makes shear and vorticity disappear with growing time.

\bigskip

{\LARGE IV. Scalar-field with torsion in an accelerating Universe}

\bigskip

\bigskip We refer to Einstein-Cartan's theory (Cartan, 1923). It was shown by
Berman (2008a), that the same model used in obtaining inflationary torsioned
solutions, can be retrieved from the Brans-Dicke case, with shear, vorticity
and lambda-term, by including a spin density, \ $S$\ \ , and a total universal
spin \ $S_{U}$\ \ , where,

\bigskip

$S_{U}=SR^{3}$ \ \ \ \ \ \ \ \ \ \ \ \ \ \ \ \ . \ \ \ \ \ \ \ \ \ \ \ \ \ \ \ \ \ \ \ \ \ \ \ \ \ \ \ \ \ \ \ \ \ \ \ \ \ \ \ \ \ \ \ \ \ \ \ \ \ \ \ \ \ \ \ \ \ \ \ \ \ \ \ (15)

\bigskip

\bigskip We refer to this last reference for the passage from Einstein's to
Jordan's frame, for spin, and for the combined field equations with torsion,
which does not affect its formula, i.e.,

\bigskip

$\bar{S}_{U}=S_{U}$ \ \ \ \ \ \ \ \ \ \ \ \ \ \ \ \ . \ \ \ \ \ \ \ \ \ \ \ \ \ \ \ \ \ \ \ \ \ \ \ \ \ \ \ \ \ \ \ \ \ \ \ \ \ \ \ \ \ \ \ \ \ \ \ \ \ \ \ \ \ \ \ \ \ \ \ \ \ \ \ (16)

\bigskip

It must be clear that the spin density term must be included in the resulting
Raychaudhuri's equation, whereby, from the relation (7) we obtain the new one
by making the substitution below,

\bigskip

\bigskip$\Lambda\longrightarrow\left[  \Lambda+128\pi^{2}S^{2}\right]  $
\ \ \ \ \ \ \ \ \ \ \ \ \ . \ \ \ \ \ \ \ \ \ \ \ \ \ \ \ \ \ \ \ \ \ \ \ \ \ \ \ \ \ \ \ \ \ \ \ \ \ \ \ \ \ \ \ \ \ \ \ \ \ (16a)

We can check, that in the given model for the BD case, we need to include an
Einstein-Cartan's theory with,

\bigskip

\bigskip$S^{2}=s_{0}^{2}t^{\gamma}$ \ \ \ \ \ \ \ \ \ \ \ \ \ \ \ \ , \ \ \ \ \ \ \ \ \ \ \ \ \ \ \ \ \ \ \ \ \ \ \ \ \ \ \ \ \ \ \ \ \ \ \ \ \ \ \ \ \ \ \ \ \ \ \ \ \ \ \ \ \ \ \ \ \ \ \ \ \ \ \ (17)

\bigskip

where \ $\gamma$\ \ \ has the value of Section III\ (\ $\gamma=-2$\ ).

\bigskip

Therefore, we obtain, \ 

$\bigskip$

$\bar{S}_{U}=s_{0}\left(  mD\right)  ^{\frac{3}{m}}t^{\frac{3}{m}-1}$ \ \ \ \ \ \ \ \ \ \ .\ \ \ \ \ \ \ \ \ \ \ \ \ \ \ \ \ \ \ \ \ \ \ \ \ \ \ \ \ \ \ \ \ \ \ \ \ \ \ \ \ \ \ \ \ \ \ \ \ \ \ \ \ \ (18)

\bigskip

This closes the calculation.

\ \ 

\bigskip

{\LARGE \bigskip V. Conclusions}

\bigskip The "no-hair" property of the otherwise perfect fluid, has been
proved for the present model. As to the spin of the Universe, \ we have two comments:

\bigskip

-- \textbf{FIRST}) the form of the spin is of the type, \ 

\bigskip

$\bar{S}_{U}$\ $\propto R^{3}t^{-1}$\ \ \ \ \ \ .

\bigskip

In the Machian limit, the scale-factor is proportional to the age of the
Universe. Then, the Machian contact with the above formula, is fulfilled by
considering, \ 

\bigskip

\ $\bar{S}_{U}$\ $\propto R^{2}$\ \ \ \ \ \ .

\bigskip

\bigskip The papers by Berman cited above, contain this Machian conclusion.

\bigskip

\textbf{-- SECOND}) though the spin grows with time, Berman (2008; 2008a;
2008b) has shown that the angular speed decreases with \ \ $R^{-1}$\ \ (de
Sabbata and Gasperini, 1985).

\bigskip

{\Large Acknowledgements}

\bigskip

The author also thanks his intellectual mentors, Fernando de Mello Gomide and
M. M. Som, and also to Marcelo Fermann Guimar\~{a}es, Nelson Suga, Mauro
Tonasse, Antonio F. da F. Teixeira, and for the encouragement by Albert, Paula
and Geni.

\bigskip

\bigskip{\Large References}

\bigskip\bigskip Berman, M.S. (1983) - Nuovo Cimento, \textbf{74B}, 182-6.

\bigskip\bigskip Berman, M.S. (2007) - \textit{Introduction to General
Relativistic and Scalar Tensor Cosmologies}, Nova Science, New York.

Berman,M.S. (2007a) - \textit{Introduction to General Relativity and the
Cosmological Constant Problem}, Nova Science, New York.

Berman,M.S. (2007b) - \textit{The Pioneer Anomaly and a Machian Universe,
}Astrophysics and Space Science, \textbf{312}, 275.

Berman,M.S. (2007c) - \textit{Shear and Vorticity in Inflationary Brans-Dicke
Cosmology with Lambda-Term, }Astrophysics and Space Science, \textbf{310, }205.

Berman,M.S. (2008) - \textit{A Primer in Black Holes, Mach's Principle and
Gravitational Energy}, Nova Science, New York.

Berman,M.S. (2008a) - \textit{Shear and Vorticity in a Combined
Einstein-Cartan-Brans-Dicke Inflationary Lambda-Universe, }Astrophysics and
Space Science, \textbf{314, }79-82.

Berman,M.S. (2008b) - \textit{A General Relativistic Rotating Evolutionary
Universe, }Astrophysics and Space Science, \textbf{314, }319-321.

Berman,M.S. (2008c) - \textit{A General Relativistic Rotating Evolutionary
Universe - Part II, }Astrophysics and Space Science, to appear.

Berman,M.S.; Gomide, F.M. (1988) - GRG, \textbf{20,} 191-198.

\bigskip Brans, C.; Dicke, R.H. (1961) - Physical Review, \textbf{124}, 925.

Cartan, E. (1923) - Ann. Ec. Norm. Sup., \textbf{40,} 325.

Dicke, R.H. (1962) - Physical Review, \textbf{125}, 2163.

Faraoni, V. (2004) - \textit{Cosmology in Scalar-Tensor Gravity, }Kluwer, Dordrecht.

Fujii, Y.; Maeda, K-I. (2003) - \textit{The Scalar-Tensor Theory of
Graviation, }CUP, Cambridge.

\bigskip Raychaudhuri, A. K. (1979) - \textit{Theoretical Cosmology, }Oxford
University Press, Oxford.

Sabatta, V. de;\ Gasperini, M. (1985) - \textit{Introduction to Gravitation,
}World Scientific, Singapore.

\end{document}